\documentclass[12pt,preprint]{aastex}

\begin{document}

\title{Statistical features of 21 cm emission from the epoch between
  reionization and Gunn-Peterson transparency}

\author{Ping He\altaffilmark{1,2}, Jiren Liu\altaffilmark{3},
 Long-Long Feng\altaffilmark{4,1}, Hong-Guang Bi\altaffilmark{2}, and
 Li-Zhi Fang\altaffilmark{2}}

\altaffiltext{1}{National Astronomical Observatories, Chinese
Academy of Sciences, 20A Datun Road, Chaoyang, Beijing 100012,
China}
\altaffiltext{2}{Department of Physics, University of
Arizona, Tucson, AZ 85721}
\altaffiltext{3}{Center for
Astrophysics, University of Science and Technology of China, Hefei
230026, China}
\altaffiltext{4}{Purple Mountain Observatory,
Nanjing 210008, China}

\begin{abstract}

We investigate the 21 cm emission from the epoch between
reionization $z_r$ and the Gunn-Peterson transparency $z_{GP}$.
According to the lognormal model of the thermal history around
reionization, hydrogen clouds in $z_r >z>z_{GP}$ are hot and a
predominant part of baryonic gas is ionized, but still opaque to
Ly$\alpha$ photons. Therefore, 21 cm emission is a distinctive
characteristic of this epoch. We show that the 21 cm emission
comes from both uncollapsed and collapsing hydrogen clouds. The
spatial distribution of the brightness temperature excess $\delta
T_b$ is highly non-Gaussian. It consists of spikes with high
$\delta T_b$ and a low $\delta T_b$ area between the spikes. The
field has the following statistical features: (1) the one-point
distributions of $\delta T_b$ are described approximately by
power-law tailed probability distribution functions; (2) the
$n$th-order moment of $\delta T_b$ is increasing much faster with
$n$ than that of a Gaussian field, but slower than that of a
lognormal field; (3) the scale-scale correlation of $\delta T_b$
field is significant for all scales larger than the Jeans length
of the gas. These features would be useful for distinguishing the
21 cm emission of the early clustering from the noise of
foreground contamination.

\end{abstract}


\section{Introduction}

The universe becomes transparent to Ly$\alpha$ photons (the
Gunn-Peterson transparency) at redshift $z_{GP} \simeq 6.5$ (Fan
et al. 2002). Yet the polarization of the cosmic microwave
background (CMB) indicates that the Compton scattering optical
depth to the CMB is as high as $\tau_e \simeq 0.17 \pm 0.04$ and
that the reionization of the universe probably occurred at $11
<z_r <30$ (Kogut et al. 2003). Most theoretical attempts to
discriminate between the redshifts $z_{GP}$ and $z_r$ concluded
that the history from first-generation star formation to the
Gunn-Peterson transparency lasts a long time (Cen 2003; Haiman \&
Holder 2003; Whyithe \& Loeb 2003; Holder et al. 2003; Onken \&
Miralda-Escud\'e 2003).

This problem has recently been studied by using the lognormal (LN)
formalism (Bi et al. 2003; Liu et al. 2004). Considering that the
Jeans length changed greatly before and after the reionization, a
self-consistent calculation with the LN model yields a general
picture that the redshift $z_r$ should be significantly larger
than $z_{GP}$, where $z_r$ means the era that ionized regions are
fully overlapped. The thermal history of the universe around the
reionization can be roughly divided into three epochs: (1) the
cold dark age $z>z_r$, (2) the hot dark age $z_r
> z> z_{GP}$, and (3) the bright age $z<z_{GP}$. With the cosmological
parameters given by WMAP and COBE, the hot dark age lasts about
$z_r - z_{GP}\simeq 10$.

In the epoch $z_r > z> z_{GP}$, the fraction of ionized hydrogen
$f_{HII}= n_{HII}/(n_{HI}+n_{HII})$, where $n_{HI}$ and $n_{HII}$
are the number densities of neutral and ionized hydrogen atoms, is
already significant, while the mass fraction of neutral hydrogen
$f_{HI}= n_{HI}/(n_{HI}+n_{HII})$ is still much greater than
$10^{-3}$, so that the universe is opaque to the Ly$\alpha$
photons. Moreover, the temperature of hydrogen gas in the period
$z_r > z > z_{GP}$ is higher than that of the CMB. Therefore, in
this epoch, the 21 cm emission given by the hyperfine structure of
the ground state $1^2S_{1/2}$ of neutral hydrogen would be
significant. This motivates us to investigate in this paper the
statistical properties of the redshifted 21 cm emission from the
epoch $z_r > z> z_{GP}$.

Many calculations on the redshifted 21 cm radiation from the early
universe have been done based on various models of the thermal
history around reionization (Ciardi \& Madau 2003; Furlanetto \&
Loeb 2002, 2004;  Iliev et al. 2002, 2003; Tozzi et al. 2000;
Zaldarriaga et al. 2004; Furlanetto et al. 2004). These works have
generally focused on the power spectrum of the fluctuations of the
21 cm emission. In this paper, besides the power spectrum, we
address the higher order statistics and non-Gaussianity of the
spatial distribution of the 21 cm emission. Since the 21 cm
emission at high redshifts is relevant to the earliest
gravitational clustering and star formation of the universe, the
field of the 21 cm emission must be non-Gaussian. It would
probably be the earliest non-Gaussian field if the initial
perturbations of cosmic mass density field are Gaussian.

The non-Gaussianity of 21 cm emission field is not only
theoretically important but also observationally useful. Radio
observation would be able to detect the cosmic 21 cm emissions if
its flux is of the order of $\simeq 0.1$ mJy or higher and angular
scale is of the order of arcminutes (Morales and Hewitt 2004; Pen
et al. 2004). However, it is a challenge to identify the
redshifted 21 cm emission from the recombination epoch, since the
noise from the artificial radio interference in the VHF band is
serious. The non-Gaussian statistical features would be helpful to
draw the information from the noisy observations.

This paper is outlined as follows. \S 2 briefly describes the
basic results of the reionization and thermal history of hydrogen
clouds in the LN model. \S 3 presents the simulations of the
baryon field and the 21 cm emission from neutral hydrogen in the
epoch of $z_r$ to $z_{GP}$. The basic properties of the 21 cm
emission from diffused intergalactic medium (IGM) and collapsed
halos are also addressed. \S 4 presents a statistical analysis of
the $\delta T_b$ fields, including the first-, second- and
high-order statistics. Finally \S 5 gives the discussion and
conclusion.

\section{Evolution of hydrogen clouds around reionization}

\subsection{Hydrogen gas distribution in LN model}

The LN model assumes that the mass density distribution $\rho({\bf
x})$ of hydrogen gas at redshift $z$ is given by an exponential
mapping of the linear distribution of the dark matter mass density
contrast $\delta_0 ({\bf x})$, which is smoothed on the Jeans
length scale of the gas at the redshift $z$ (Bi 1993), i.e.,
\begin{equation}
\rho({\bf x}) = \bar{\rho}_0\exp[\delta_0 ({\bf x}) -  \sigma_0^2/2],
\end{equation}
where $\bar{\rho}_0$ is the mean density and $\sigma_0=\langle
\delta_0 ^2 \rangle ^{1/2}$ is the variance of the Gaussian mass
field $\delta_0 ({\bf x})$ on the scale of the Jeans length of the
gas. The probability distribution function (PDF) of the density
field $\rho({\bf x})$ is, then, lognormal:
\begin{equation}
p(\rho/\bar{\rho})=\frac{1}{(\rho/\bar{\rho}) \sigma_0\sqrt{2\pi}}
 \exp\left [ -\frac{1}{2}
  \left(\frac{\ln (\rho/\bar{\rho}) + \sigma_0^2/2}{\sigma_0}\right )^2
   \right ], \hspace{3mm} \rho \geq 0.
\end{equation}
The dynamical arguments of the exponential mapping and LN PDF of
both the dark matter and hydrogen clouds have been studied by,
e.g., Coles \& Jones (1991), Jones (1999), and Szapudi \& Kaiser
(2003).

The first characteristic of the LN model is that the clustering of
hydrogen clouds is mainly dependent on the variance $\sigma_0$ of
the linear fluctuations on the scale of the Jeans length, while
the clustering of dark matter basically is independent of the
properties of gas. Therefore, the LN model is effective in
describing the effect of the thermal status of hydrogen gas on the
density and velocity field of the baryonic gas. This has been
shown in the successful application of modelling IGM Ly$\alpha$
forests of QSO's absorption spectra in redshift range $2 \leq z
\leq 5$ (Bi 1993; Bi \& Davidsen 1997; Feng \& Fang 2000).

The second characteristic is that, compared to a Gaussian or a
normal distribution, the PDF of an LN model described by equation
(2) has a prolonged tail at the high-density end and that the
probability of high-density events $\rho/\bar{\rho}\gg 1$ is very
sensitive to $\sigma_0$. This property has already been found to
be useful in explaining the intermittence of the transmitted flux
of QSO's Ly$\alpha$ absorption spectrum (Jamkhedkar et al. 2000,
2003; Pando et al. 2002). The long tail can directly be seen with
(1) the cumulative mass fraction $M(>\rho/\bar{\rho})$, which is
the fraction of mass in regions having mass density larger than a
given $\rho$, and (2) the volume filling factor
$V(>\rho/\bar{\rho})$, which is the fraction of volume with
density larger than a given $\rho$. They are
\begin{equation}
M[>(\rho/\bar{\rho})] = \int ^{\infty}_{\rho/\bar{\rho}} x p(x) dx
= \frac{1}{2} {\rm erfc} \left (\frac{\ln
(\rho/\bar{\rho})}{\sqrt{2} \sigma_0}-
\frac{\sigma_0}{2\sqrt{2}}\right ),
\end{equation}
\begin{equation}
V(>\rho/\bar{\rho}) = \int ^{\infty}_{\rho/\bar{\rho}} p(x) dx
= \frac{1}{2} {\rm erfc}\left (\frac{\sigma_0}{2\sqrt{2}} +
 \frac{\ln (\rho/\bar{\rho})}
{\sqrt{2} \sigma_0} \right ).
\end{equation}
Therefore, even when $\sigma_0\simeq 1$, the mass fraction of
high-density events is already significant and their volume
filling factor is very small.

\subsection{The clustering of hydrogen clouds at redshifts $z_r$ and
   $z_{GP}$}

In the context of the LN model, the evolution of hydrogen clouds
is mainly dependent on the Jeans length of hydrogen gas. If
hydrogen gas is uniformly distributed, its Jeans length is given
by $\lambda_J\equiv v_s(\pi/G\rho)^{1/2}$, where $v_s$ is the
sound speed of the gas. For cosmological study, it is convenient
to use a comoving scale $x_J =\lambda_J/2\pi= (1/H_0)[2\gamma
k_BT_m/3\mu m_p\Omega(1+z)]^{1/2}$, where $T_m$ is the mean
temperature, $\mu$ the molecular weight of the gas, $\Omega$ is
the cosmological density parameter of total mass, and $\gamma$ is
the ratio of specific heat.

Primordial baryons, created at the time of nucleosynthesis,
recombine with electrons to become neutral gas at $z \simeq 1000$.
Before $z \simeq 200$, the residual ionization of the cosmic gas
keeps its temperature locked to the CMB temperature (Peebles
1993). After $z \simeq 200$, the gas cools down adiabatically
because of the expansion of the universe. Assuming the adiabatic
index $\gamma=5/3$, i.e., hydrogen temperature $T\propto
\rho^{2/3}$, the comoving Jeans length evolves as $\simeq 0.1
\times (1+z)^{1/2} \ h^{-1}$ kpc. During reionization, the gas is
heated by UV ionizing photons from a low temperature to $\sim 1.3
\times 10^4$K. This leads to the increase of $x_J$ at $z_r$.

Assuming $z_r\simeq 18$ (Kogut et al. 2003; Liu et al. 2004), the
evolution of the Jeans length is sketched in Figure 1, which gives
the relation between $x_J$ and the cosmic scale factor $1/(1+z)$.
The sudden increase of $x_J$ at $z_r$ shown in Figure 1 is
certainly unrealistic, which comes from the assumption that the
gas temperature increases instantly by a factor of about $10^{4}$
at $z_r$. More realistically, the sharp changes of $x_J$ at $z_r$
and $z_{GP}$ should be replaced by a softened transition.
Nevertheless, the physical status of the gas before and after
$z_r$ is properly sketched in Figure 1.

Corresponding to the evolution of $x_J$ at $z_r$, the variance
$\sigma_0$, which is the ${\it rms}$ mass fluctuation on the Jeans
length $(\delta \rho/\rho)_{x_J} \propto k^3P(k)|_{k=1/x_J}$, also
undergoes a zigzag evolution around $z_r$. Figure 2 shows
$\sigma_0$ versus $1/(1+z)$ calculated with the $x_J$ of Figure 1.
To calculate $\sigma_0$, the power spectrum $P(k)$ of the linear
mass density perturbations is taken to be the spectrum of the
low-density flat cold dark matter model (LCDM), which is specified
by the density parameter $\Omega_0=0.3$, the cosmological constant
$\Omega_{\Lambda}=0.7$, and the Hubble constant $h=0.7$. The
linear power spectrum $P(k)$ is given by the fitting formula given
by Eisenstein \& Hu (1999). The increase of $\sigma_0$ with
$1/(1+z)$ shown in Figure 2 is due to the gravitational linear
growth. The variance at small redshifts $(z <7)$ shown in Figure 2
actually is the same as that given by Bi \& Davidsen (1997). The
zigzagged feature at $z\simeq3.3$ of Figure 2 is due to HeII
reionization (Theuns et al. 2002).

An important feature of $\sigma_0$ shown in Figure 2 is that the
curve of $\sigma_0$ from $A'$ to $B'$ almost repeats that from $A$
to $B$. Thus, the evolution of the fraction of mass [eq.(3)] in
the period from $A'$ to $B'$ is similar to that from $A$ to $B$.
The period from $A$ to $B$ corresponds to an evolution of
weak-to-strong clustering of the baryonic gas, and the period from
$A'$ to $B'$ also corresponds to an evolution of weak-to-strong
clustering of baryonic gas. Therefore, the formation rate of
collapsed hydrogen clouds at $A'$ is significantly lower than that
at $B$. This leads to a suppression of clustering and star
formation just after $z_r$.

The dynamical picture of the suppression of clustering can be
explained by the following negative feedback mechanism. During the
reionization $z_r$, hydrogen clouds are heated by the ionizing
photons so that the mean temperature of hydrogen clouds increases
by a few orders. All the irregularities originally formed in cold
gas on scales smaller than the new Jeans length would be smoothed
out by the heated gas, and the variance of the fluctuations of
hydrogen cloud distribution drops back below unity. Although dark
matter halos continuously collapse before and after $z_r$,
hydrogen clouds on scales less than the new Jeans length will stop
collapsing. As a result, star formation will be slowed down or may
be even halted in these halos after $z_r$. The level of the star
formation suppression is adjusted by the following mechanism. Once
the star formation rate declines, the UV background produced by
the star formation also becomes lower, and so do the temperature
and entropy states of hydrogen gas, which make it easy for further
star formation. On the contrary, if the star formation rate
increases, the UV background produced by the star formation also
increases, and then more hydrogen gas will be heated to higher
temperature and higher entropy states. This finally yields more
suppression in the rate of star formation.

When the variance $\sigma_0$ grows again to unity around $B'$
because of the increase of the potential wells of dark matter,
there is once again enough to have collapsed hydrogen clouds, star
formation, and ionizing photons. This finally gives rise to the
Gunn-Peterson transparency. Therefore, the long-lasting period
from $z_r$ to $z_{GP}$ is due to the zigzag of $\sigma_0$ around
reionization $z_r$. The difference $z_r-z_{GP}$ is found to be
equal to $\simeq 10$. It is weakly dependent on, or stable with
respect to, the following parameters: (1) the threshold of the
collapsed hydrogen clouds $\rho/\bar{\rho}$ and (2) the mean
ionizing photons released by each baryon in the collapsed objects
(Liu et al. 2004). This stability is due to the above-mentioned
adjustment mechanism. In this sense, the redshift of reionization
$z_r \simeq 18$ actually is determined by the redshift of the
Gunn-Peterson transparency $z_{GP}$ (Liu et al. 2004). This result
is also in good agreement with the observed optical depth $\tau_e$
to the CMB photons.

\subsection{Temperature of hydrogen clouds in the epoch of
   $z_r$ to $z_{pg}$}

According to the above-discussed scenario, in the epoch $z>z_r$,
hydrogen gas is cold and neutral. It may contain some individual
ionized spheres around the first-generation stars. Hydrogen clouds
generally are opaque to Ly$\alpha$ photons. When $z$ approaches
$z_r$, the increasing individual ionized regions overlap. At
$z_r$, the individual ionized regions fully overlap in the sense
that the universe becomes transparent to soft X-ray photons. Thus,
after $z_r$, hydrogen clouds are heated by soft X-ray photons and
the star formation is suppressed. The fraction of ionized hydrogen
($HII$) will no longer increase, or will even decrease after
$z_r$. Consequently, hydrogen clouds are still opaque to
Ly$\alpha$ photons.

It has been shown that if the photoionization heating is the major
heating process of the hydrogen clouds, the mean gas temperature
is always of the order of $\sim 10^4$ K, weakly dependent on the
intensity of heating photons (Black 1981; Ostriker \& Ikeuchi
1983). This result is further supported by hydrodynamic
simulations of baryonic gas in the early universe (e.g., Hui \&
Haiman 2003). They showed that the mean gas temperature
asymptotically approaches $10^3-10^4$ K. This temperature is not
determined by the intensity of the heating photons, but is
completely given by the shape (or hardness) of the spectrum of the
heating photons. The mean temperature is higher for harder photon
spectra, such as soft X-rays. Thus, although star formation is
slowing down after $z_r$, the mean temperature will still be in
the range $10^3-10^4$ K. Therefore, it is reasonable to assume
that in the epoch of $z_r$ to $z_{GP}$, the mean gas temperature
can be described as $T_m \simeq T_0[(1+z)/(1+z_r)]^{\eta}$, where
$T_0\simeq 10^4$K and the index $\eta$ is of the order of 1. In
Figure 3, we plot the mean temperature evolution for the two cases
of $(T_0, \eta)=(0.8\times 10^4K, 2)$ and $(1.3\times 10^4 K, 2)$.
In Figures 1 and 2 we also show, respectively, the temperature
effect on the Jeans length $x_J$ and variance $\sigma_0$ in the
epoch $z_r > z > z_{GP}$.

If hydrogen gas is polytropic, gas temperature at the point with
mass density $\rho$ is given by
$T=T_m(\rho/\bar{\rho})^{\gamma-1}$. In this way, we can calculate
the thermal status of hydrogen from the density distribution. In
the LN model, the relation $T-\rho$ and the index $\gamma$
actually are given by fitting the hydrodynamical simulation (Bi \&
Davidsen 1997). In this case, shock heating is partially
considered. However, if gravitational shocks are strong, the
temperature of hydrogen gas is basically multiphased, and one
cannot describe the $T-\rho$ relation by a single equation. To
estimate the effect of strong shock, we have to replace the
temperature-density relation $T=T_m(\rho/\bar{\rho})^{\gamma-1}$
with $P(T; \rho)$, the PDF of temperature at a given $\rho$. Such
a PDF has been recently derived by using a hydrodynamic simulation
code that is very effective in capturing shocks (He et al. 2004).
He et al. (2004) found that strong shock events are very rare at
$z> 4$, i.e., that the events leading to the deviation from the
relation $T=T_m(\rho/\bar{\rho})^{\gamma-1}$ are negligible at
$z>4$.

\section{Samples from the LN simulation}

\subsection{Simulation of the hydrogen gas distribution}

We produce simulation samples of spatial distribution of hydrogen
gas $\rho({\bf x})$ by the LN model. In order to quickly grasp the
features of these distributions, we only concentrate on samples of
one-dimensional distribution. The details of the simulation
procedures have been given in Bi \& Davidsen (1997) and Bi et al.
(2003). A brief description is as follows.

We first generate the one-dimensional density and velocity
distributions in Fourier space, $\delta_0(k,z)$ and $v(k,z)$,
which are two Gaussian random fields. Both $\delta_0(k, z)$ and
$v(k, z)$ are given by the power spectrum, $P_0(k)$ as follows (Bi
1993; Bi et al. 1995)
\begin{equation}
\delta _0 (k, z) = D(z) (u(k) + w(k)),
\end{equation}
\begin{equation}
v(k, z) = F(z) \frac{H_0}{c} ik \alpha (k) w(k),
\end{equation}
where $D(z)$ and $F(z)$ are the linear growth factors for the
$\delta _0( x)$ and $v(x)$ fields at redshift $z$. The $w(k)$ and
$u(k)$ fields are Gaussian with the power spectra given by
\begin{equation}
P_w(k) = \alpha ^{-1} \int ^\infty _k P_{0}(q) 2\pi q^{-1} dq,
\end{equation}
\begin{equation}
P_u(k) = \int ^\infty _k P_{0}(q) 2\pi q dq - P_w(k),
\end{equation}
where $P_0(k)$ is the power spectrum of the three-dimensional
field $\delta_0({\bf x})$. The functions $\alpha(k)$ in equation
(7) is defined by
\begin{equation}
\alpha(k) = \frac{\int ^\infty _k P_{0}(q) q^{-3} dq}
            {\int ^\infty _k P_{0}(q) q^{-1} dq}.
\end{equation}
The power spectrum $P_0(k)$ is taken to be (Bi \& Davidsen 1997;
Bi et al. 2003)
\begin{equation}
P_{0}(k) = \frac{P_{dm}(k)}{(1+x_J ^2 k^2)^2}.
\end{equation}
Obviously, $P_{0}(k)$ is also a function of $z$ via the redshift
dependence of $x_J(z)$. Thus, for a given $P_{DM}(k)$ and $x_J$,
one can produce the distributions $\delta_0(k, z)$ at the grid
points $k_i$, $i=1, 2, ..., N$ in one-dimensional Fourier space.
The spatial distributions $\delta_0(x, z)$ can be obtained by
using the fast Fourier transform. Since the velocity follows the
linear evolution longer than the density, we can use the linear
$v(k,z)$ and its Fourier counterpart as good approximations of the
velocity field at high redshifts. In Figure 4 we plot a typical
realization of the gas mass density fields at redshifts 7, 10, and
15. The simulation size shown in Figure 4 is 50 $h^{-1}$ Mpc in
comoving space. The total number of pixels is $2^{14}=16,384$. The
pixel size is then $\simeq 3 \ h^{-1}$ kpc, which is less than the
smallest scale of $x_J$ at $z<z_r$. Therefore, the samples are
qualified for studying the non-Gaussian features of the hydrogen
clouds on small scales.

As analyzed in \S 2.3, the effect of strong gravitational shock at
high redshifts is negligible and the temperature of hydrogen gas
can be calculated by the relation $T
=T_m(\rho/\bar{\rho})^{\gamma-1}$. We then have the temperature
field of the hydrogen clouds. With the mass density and
temperature fields, one can find the neutral hydrogen mass density
fields with the photoionization equilibrium equation as
\begin{equation}
f_{HI} = \frac{\alpha (T)}{\alpha (T) + \Gamma (T) + J/n_e},
\end{equation}
where $\alpha(T)$ is the recombination rate, $\Gamma (T)$ is the
collisional ionization rate, $n_e$ is the number density of
electrons, and $J$ the rate of photoionization of hydrogen (Black
1981). Although equation (11) is reasonable for determining
$f_{HI}$ of the IGM at $z<z_{GP}$, there are two points should be
addressed when applying it to the epoch $z_r - z_{GP}$.

First, equation (11) is applicable only if the hydrogen clouds are
optically thin for photons of $J$. However, hydrogen clouds of
$z_r> z > z_{GP}$ are optically thick for Ly$\alpha$ photons and
may also be so for photons of $\sim 13.6$ eV. Therefore, in our
case, the flux $J$ in equation (11) is not for Ly$\alpha$, but for
soft X-ray photons. As discussed in \S 2.3, in the epoch $z_r > z
> z_{GP}$, hydrogen clouds are transparent to soft X-ray photons,
which are the major sources of photoionization heating. This
picture is consistent with the fact that the photon spectra from
the first-generation (massive) stars are generally hard.

Second, in the case of $z<z_{GP}$, the parameter $J$ is determined
by fitting it with the mean transmitted flux of Ly$\alpha$ photons
(e.g., Choudhury et al. 2001). Obviously, no such observations can
be used to constrain the parameter $J$ in the epoch of $z_r > z >
z_{GP}$. However, $J/n_e$ can be adjusted by fitting it to the
mean fractions of $HI$ or $HII$, which are required by the optical
depth $\tau_e$. Using this method, we can produce samples of
$f_{HI}$ field. Both the models of $(T_0, \eta)$=$(0.8 \times 10^4
K, 2)$ and $(1.3 \times 10^4 K, 2)$ are consistent with the
optical depth $\tau_e$ from the WMAP.

\subsection{Simulation of the 21 cm emission of hydrogen clouds}

With the sample of gas mass density, temperature, and $f_{HI}$
distributions, we can calculate the 21 cm ($\nu_0=$ 1420 MHz)
emission of neutral hydrogen (HI). This emission at redshift $z$
is determined by the difference between the spin temperature of
neutral hydrogen, $T_s(z)$, and the temperature of the cosmic
microwave background, $T_{CMB}(z)=2.73(1+z)$ K. There are two
mechanisms leading to $T_s(z) > T_{CMB}$, collision and radiation
background, which gives (Field 1958, 1959)
\begin{equation}
T_s= \frac{T_{CMB}+y_cT_c+ y_LT_L}{1+y_c+y_L},
\end{equation}
where $T_c$ is the temperature of hydrogen gas, $T_L$ is the
temperature of Ly$\alpha$ photons, and $y_c$ and $y_L$ are the
collision and radiative excitation efficiencies, respectively.
Since hydrogen clouds are optically thick to the Ly$\alpha$
photons, it is reasonable to assume that Ly$\alpha$ photons are in
approximate thermal equilibrium with the gas. Thus, one can take
$T_L$ to be the gas temperature.

Considering self-absorption, the brightness temperature of the 21
cm radiation at redshift $z$ is determined from the radiation
transfer equation as
\begin{equation}
 T_b(z)= T_{cmb}(z)e^{-\tau(z)}
    +\int^{\tau(z)}_{0}T_s(z')e^{-\tau(z')}d\tau(z'),
\end{equation}
where the first term on the right-hand side is from the CMB and
the second term is from neutral hydrogen; $\tau(z)$ is the optical
depth of the 21 cm absorption. When $T_s$ is much larger than
$T_{*}=h\nu_0/k_B=0.06$ K, $\tau(z)$ is given by (Wild 1952)
\begin{equation}
\tau(z) = \frac{3hc^3A_{10}}{32\pi\nu_0^2 k_B}
\int_0^{z_r}\frac{n_{HI}(z')}{T_s(z')H(z')}F(z, z')dz',
\end{equation}
where factor $F(z, z')$ is the normalized line profile. For
Doppler broadening, we have
\begin{equation}
F(z, z') = \frac{1}{\sqrt{\pi}b(1+z)}
   e^{-\left (\frac{z'-z}{b(1+z)}\right )^2},
\end{equation}
where $b= (2k_BT/mc^2)^{1/2}$. For gas with temperature $\leq10^5$
K, we have $b \leq 10^{-4}$, and therefore the integral of
equation (14) lasts for a very narrow range $z\pm \Delta z$ and
$\Delta z \leq 10^{-4}(1+z)$. Thus, the velocity distortion can be
ignored in the first stage.

If $n_{HI}(z)$ and $T_s(z)$ are not strong functions of $z$,
equation (14) gives approximately (Field 1959)
\begin{equation}
\tau(z) = \frac{3hc^3A_{10}n_{HI}(z)}{32\pi\nu_0^2 k_BT_s(z)H(z)}.
\end{equation}
Therefore the fields $n_{HI}(z)$ and $T_s(z)$ in equation (16)
are, respectively, the HI number density and spin temperature
$T_s(z)$ smoothed by the window function equation (15). Thus,
equation (13) yields approximately $T_b(z)=
T_{CMB}(z)e^{-\tau(z)}+T_s(z)(1-e^{-\tau(z)})$, and therefore the
observed brightness temperature excess at the redshifted frequency
$\nu=\nu_0/(1+z)$ is
\begin{equation}
\delta T_b(\nu) =[T_s(z) -T_{cmb}(z)]\frac{1-e^{-\tau(z)}}{1+z}.
\end{equation}

We show in Figure 5 a typical realization of the fields of $T_b$,
$\tau(z)$, and $\delta T_b$ at $z$=7 in the temperature model
$(T_0, \eta)=(1.3 \times 10^4 K, 2)$. Figure 5 shows that the
field of the brightness temperature excess, $\delta T_b$, is
highly nonuniform. At most places, $\delta T_b$ is actually zero
but contains many high spikes, or patchy structures. The average
of $\delta T_b$ is only $\simeq 2.3 mK$, while the spikes can be
as high as 20 - 40 mK. The rms of the brightness temperature
fluctuations, $\langle \delta T_b^2\rangle^{1/2}$, is $\simeq$ 7
mK. For $(T_0, \eta)=(0.8 \times 10^4 K, 2)$, we have $\langle
\delta T_b^2\rangle^{1/2}= 9.5$ mK. It has been argued that the
brightness temperature excess would be too small to observe when
considering the foreground noise contamination. However, the
nontrivial features of the non-Gaussianity might help identify the
redshifted 21 cm emission from the contaminated data.

The non-Gaussianity of the $\delta T_b$ field can also be seen
with $(d\langle\delta T_b^2\rangle^{1/2}/d\rho) d\rho$, which is
the fraction of $\langle\delta T_b^2\rangle^{1/2}$ given by
regions with hydrogen gas density from $\rho$ to $\rho$+$d\rho$.
Figure 6 plots the integrated fraction function $\langle\delta
T_b^2\rangle^{1/2}(> \rho)=\int_{\rho}^{\infty} d\langle\delta
T_b^2\rangle^{1/2}$. It shows that about one-third of
$\langle\delta T_b^2\rangle^{1/2}$ is given by regions with
density $2 < \rho <6$ and that the other two-thirds are from
regions of $\rho \geq 6$. Because the $\rho \geq 6$ regions
correspond to the collapsed/collapsing hydrogen clouds (Bi et al.
2003; Liu et al. 2004), Figure 6 shows that both diffused and
collapsed hydrogen clouds of the mass field have comparable
contributions to the brightness temperature fluctuations of the
redshifted 21 cm emission. Therefore, the non-Gaussianity caused
by the quasi-linear and the nonlinear clustering of hydrogen
clouds should be significant.

\section{Statistical properties of 21-cm emissions}

\subsection{DWT variables of $\delta T_b$ field}

In order to effectively describe statistical features of $\delta
T_b$ field, we decompose the $\delta T_b$ field by the discrete
wavelet transform (DWT; Fang \& Feng 2000; Fang \& Thews 1998).
Because the DWT modes are spatially localized, each dimension can
be treated separately, and hence it is easy to match the
simulation with the physical condition of the sampling. The
three-dimensional power spectrum can be effectively recovered by
DWT from the data in an area with ``poor" geometry. Therefore,
with the DWT, one can effectively compare and test the predicted
statistical properties from one-dimensional or two-dimensional
simulation with various observations (e.g., Guo et al. 2004 and
references therein).

In the DWT scheme, there are two sets of bases given by (1)
scaling functions $\phi_{j,l}({x})=\langle {x}|{j,l}\rangle_s$ and
(2) wavelets $\psi_{j,l}({x})=\langle {x}|{j,l}\rangle_w$, where
$j=0,1,..$ and $l=0, 1...2^{j}-1$. In one-dimensional space with
size $L$, the scaling function $\phi_{j,l}( x)$ is localized in
physical space $lL/2^{j}  < x \leq (l+1)L/2^{j}$ while wavelet
$\psi_{j,l}$ is localized in both physical space $lL/2^{j}  < x
\leq (l+1)L/2^{j}$ and Fourier space $\pi 2^{j}/L < |k| <(3/2)2\pi
2^{j}/L$.

The scaling functions are orthonormal with respect to the index
$l$ as
\begin{equation}
\ _{s}\langle {j,l}| {j,l'}\rangle_{s}
\equiv\int \phi_{ j,l}({\bf x})\phi_{ j,l'}({\bf x})d{ x}
 = \delta^K_{l,l'}.
\end{equation}
Scaling function $\phi_{j,l}( x)$ actually is a window function
for the spatial range $lL/2^{j}  < x \leq (l+1)L/2^{j}$. The
scaling function coefficient (SFC) of a field $\delta T_b(x)$ is
defined as $\epsilon_{j,l}\equiv \ _{s}\langle { j,l}| \delta T_b
\rangle =\int \delta T_b( x)\phi_{ j,l}( x)d x$, and therefore,
the mean of $\delta T_b( x)$ in the spatial range $lL/2^{j}  < x
\leq (l+1)L/2^{j}$ is
\begin{equation}
\delta T_{b,(j,l)}=\frac{\epsilon_{j,l} }
 {\ _{s}\langle {j,l}| 1 \rangle },
\end{equation}
where $| 1 \rangle$ is a uniform field with field strength equal
to unity.

The wavelets $\psi_{j,l}(x)$ are orthonormal bases with respect to
both indexes $j$ and $l$
\begin{equation}
\ _s\langle {j,l}|{j',l'}\rangle_s \equiv
\int \psi_{ j,l}( x)\psi_{j',l'}(x)dx
 =\delta_{j,j'}\delta_{l,l'}.
\end{equation}
The wavelet function coefficient (WFC) of the $\delta T_b(x)$
field is defined as
\begin{equation}
\tilde{\epsilon}_{ j,l}\equiv \ _{w}\langle { j,l}| \delta \rangle
=\int \delta T_b( x)\psi_{j,l}( x)d x.
\end{equation}
Since the set of the wavelet bases $|j,l\rangle_w$ is complete,
the $\delta T_b(x)$ field can be expressed as
\begin{equation}
\delta T_b(x)=\sum_{j}\sum_{l}\tilde{\epsilon}_{j,l}
\psi_{j,l}(x),
\end{equation}
where each $j$ runs 0, 1, 2... and $l$ runs 0, 1,...$2^{j}-1$.
Therefore, the WFCs $\tilde{\epsilon}_{j,l}$ can be used as the
variables of the field $\delta T_b$. The $\tilde{\epsilon}_{j,l}$
are the fluctuations around scales $k=2\pi n/L$ with $ n=2^{j}$
and at the physical area $l$ with size $\Delta x=L/2^{j}$.

\subsection{One-point distributions of $\delta T_{b,(j,l)}$ and
$\widetilde{\delta T_b}_{(j,l)}$}

We plot in Figure 7 the one-point distributions of $\delta
T_{b,(j,l)}$ for two temperature models $(T_0, \eta, z)=
(1.3\times 10^4 K, 2, 7)$ and $(0.8 \times 10^4 K, 2, 10)$. The
scales $j$ are taken to be 6, 8 and 10, respectively,
corresponding to comoving smoothing sizes 0.78, 0.20, and 0.05
$h^{-1}$ Mpc or angular resolutions 0$^{\prime}$.44,
0$^{\prime}$.11, and 0$^{\prime}$.028 at z=7, or 0$^{\prime}$.41,
0$^{\prime}$.10, and 0$^{\prime}$.025 at z=10. All the PDFs in
Figure 7 seem to have similar shapes. A peak at $\delta T_b \simeq
0$ is given by the low mass density ($0< \rho <2$) areas; Figure 6
shows that the areas with mass density $0< \rho <2$ do not
contribute to $\delta T_b$. Except for the peaks, the PDFs of
Figure 7 are flat in the range  $\delta T_{b,(j,l)} < 1$ mK, and
approximately of a power law at $\delta T_{b,(j,l)} > 1$ mK. The
long tail obviously is given by the spikes in the $\delta T_{b}$
field (Fig. 5). The power law tail shows a slightly smoothing
scale dependence.

We plot in Figure 8 the PDFs of the WFCs, in which
$\tilde{\epsilon}_{j,l}$ is replaced by $\widetilde{\delta
T_b}_{(j,l)}=(2^j/L)^{1/2} \tilde{\epsilon}_{j,l}$ since the
latter has the dimension of temperature. The parameters used in
Figure 8 are the same as those in Figure 7. The PDFs shown in
Figure 8 are also weakly dependent on the temperature parameters,
but the tails of the PDFs are more significantly dependent on
scale $j$ than in Figure 7. If we describe the tail by a power law
$\propto\widetilde{\delta T_b}_{(j,l)}^{-a}$, the index $a$ is
bigger with larger scales. In other words, the tails are longer
for smaller scales. This leads to the mean power $\langle
\widetilde{\delta T_b}^2_{(j,l)}\rangle$ being lower for higher
$j$ (see next subsection on power spectrum).

In either Figures 7 or 8, the mean of $\delta T_{b, (j,l)}$ or
$\widetilde{\delta T_b}_{(j,l)}$ is given by the events
corresponding to the long tail of the PDFs. That is, the maps of
$\delta T_b$ or $\widetilde{\delta T_b}_{(j,l)}$ are very
different from the Gaussian noise field.

\subsection{Second-order correlations}

The two-mode correlation of the SFCs $\langle\delta
T_{b,(j,l)}\delta T_{b,(j,l')}\rangle$ is similar to the ordinary
two-point correlation function. Figure 9 presents the correlation
function $\langle\delta T_{b,(j,l)}\delta T_{b,(j,l')}\rangle$ for
parameters $(T_0, \eta, z)=(1.3\times 10^4$ K, 2, 7) with
smoothing scales $j=12$, 10 and 8, which correspond, respectively,
to comoving sizes 0.03, 0.05, and 0.20 h$^{-1}$ Mpc, or
0$^{\prime}$.007, 0$^{\prime}$.028 and 0$^{\prime}$.11. The
two-mode correlation functions are typically of power law
$\langle\delta T_{b,(j,l)}\delta T_{b,(j,l')}\rangle \propto
r^{-\gamma}$ in the range $0.2<r<1$ $h^{-1}$ Mpc with index
$\gamma \simeq 0.6$. The central part of the correlation function,
or $r<0.2$ $h^{-1}$ Mpc, is flat because of the Jeans length
smoothing. The tail, or $r > 2 $ h$^{-1}$ Mpc, of the correlation
function approaches $\langle\delta T_{b,(j,l)}\delta
T_{b,(j,l')}\rangle = \langle\delta
T_{b,(j,l)}\rangle^2=\langle\delta T_{b}\rangle^2$, which is the
mean of $\delta T_b$.

The two-mode correlation function of the WFCs $\langle
\tilde{\epsilon}_{j,l}\tilde{\epsilon}_{j,l'}\rangle$, or $\langle
\widetilde{\delta T_b}_{,(j,l)}\widetilde{\delta
T_b}_{,(j,l')}\rangle$, generally is diagonal with respect to
$(l,l')$, especially if the initial mass perturbation is Gaussian
(Feng \& Fang 2004; Guo et al. 2004). Thus we have in general
\begin{equation}
\langle \widetilde{\delta T_b}_{,(j,l)} \widetilde{\delta
T_b}_{,(j,l')}\rangle  = P_{j}\delta^K_{l,l'},
\end{equation}
where $P_{j}$ is the DWT power spectrum of the one-dimensional
field, of which the relation with the Fourier power spectrum
$P(k)$ has been given in Fang \& Feng (2000). Figure 10 gives the
power spectrum $P_j$ for parameters $(T_0, \eta, z)=(1.3\times
10^4$ K, 2, 7) and $(0.8\times 10^4$ K, 2, 10). The power spectra
of these two cases are similar. However, if using angular scales,
they are different. For instance, although both spectra are peaked
at $j=7$, they correspond to angular scale 0$^{\prime}$.22 at
$z=7$, or 0$^{\prime}$.20 at $z=10$.

The power $\langle \widetilde{\delta T_b}^2_{,(j,l)}\rangle$
increases from $j=2$ (comoving scale 12.5 $h^{-1}$ Mpc) to $j=7$
(0.39 $h^{-1}$ Mpc). This is because the clustering is stronger on
small scales. The power $\langle \widetilde{\delta
T_b}^2_{,(j,l)}\rangle$ gradually decreases with $j$ when $j>7$
(comoving scale less than 0.39 $h^{-1}$ Mpc), this is because of
the Jeans smoothing of hydrogen gas.

\subsection{High-order moments of $\delta T_b$ field}

High-order moments are sensitive to non-Gaussianity. We use two
standard high-order moment statistics: (1) ${\langle (\delta
T_b-\langle\delta T_b\rangle)^{2n}\rangle^{1/n}}/ {\langle (\delta
T_b-\langle{\delta T_b}\rangle)^2\rangle}$ and (2) $\langle
(\delta T_{b,j}^n \rangle^{1/n}/\langle (\delta T_b)\rangle$. If
the $\delta T_b$ field is Gaussian, the high-order moment of
$\delta T_b$ satisfies the relation
\begin{equation}
\frac{\langle (\delta T_b-\langle\delta T_b\rangle)^{2n}\rangle^{1/n}}
{\langle (\delta T_b-\langle{\delta T_b}\rangle)^2\rangle}=
[(2n-1)!!]^{1/n}.
\end{equation}
On the other hand, if the $\delta T_b$ field is lognormal, we have
\begin{equation}
\frac{\langle (\delta T_b)^n \rangle^{1/n}}{
\langle \delta T_b\rangle }= \exp [(1/2)(n-1)\sigma^2],
\end{equation}
where $\sigma$ is the variance of $\ln(\delta T_b)$.

We plot in Figure 11 the high-order moments ${\langle (\delta
T_b-\langle\delta T_b\rangle)^{2n}\rangle^{1/n}}/ {\langle (\delta
T_b-\langle{\delta T_b}\rangle)^2\rangle}$ and $\langle (\delta
T_{b,j}^n \rangle^{1/n}/\langle (\delta T_b)\rangle$ for the
parameters $(T_0, \eta, z)= (1.3\times 10^4 K, 2, 7)$ and $(0.8
\times 10^4 K, 2, 10)$. As a comparison, we also plot the
corresponding Gaussian and LN moments in Figures 11a and 11b,
respectively. Figure 11 shows that the $\delta T_b$ field is
neither Gaussian nor lognormal. The high-order moments are always
significantly higher than a Gaussian field for $n>1$, but always
less than an LN field. In both temperature models, the two
high-order moments are always quickly increasing with $n$ when
$n\leq 5$, and then slowly increasing when $n>5$. Therefore, the
non-Gaussianity of the $\delta T_b$ field is less than the mass
density field of hydrogen clouds. This is because high-density
areas correspond to lower $f_{HI}$ regions. The non-Gaussianity is
weakened when transferring the mass field to the $f_{HI}$ field by
equation (11). Moreover, equation (12) leads to further weakening
of the non-Gaussianity given by high-density peaks. Therefore,
although its PDFs (Fig. 7) are long tailed, the non-Gaussianity of
the $\delta T_b$ field is not as strong as that of an LN field.

High-order moments are more sensitive to the tails of $\delta T_b$
PDFs, and therefore, the statistics ${\langle (\delta
T_b-\langle\delta T_b\rangle)^{2n}\rangle^{1/n}}/ {\langle (\delta
T_b-\langle{\delta T_b}\rangle)^2\rangle}$ and $\langle (\delta
T_{b,j}^n \rangle^{1/n}/\langle (\delta T_b)\rangle$ are less
dependent on the lower limit of observable $\delta T_b$. If we use
only the observable events with $\delta T_b\geq 1$ mK, the
high-order moments given by the truncated PDFs [Fig. 7] will not
be changed very much. Moreover, the statistics ${\langle (\delta
T_b-\langle\delta T_b\rangle)^{2n}\rangle^{1/n}}/ {\langle (\delta
T_b-\langle{\delta T_b}\rangle)^2\rangle}$ and $\langle (\delta
T_{b,j}^n \rangle^{1/n}/\langle (\delta T_b)\rangle$ are defined
by ratios of $\delta T_b$. Hence, they are less dependent on mean
$f_{HI}$. The statistical behavior shown in Figure 11 is weakly
dependent on ionizing photon parameter $J$ in equation (11).

\subsection{Scale-scale correlations of the WFCs}

The so-called scale-scale correlations of a random field measures
the correlation between the fluctuations on different scales
(Pando et al. 1998). For $\delta T_b$ field, this statistics is
defined as
\begin{equation}
C^{p,p}_{j} = \frac{\langle \tilde{\epsilon}^p_{j,{[l/2]}}
  \tilde{\epsilon}^p_{j+1,l}\rangle}
{\langle \tilde{\epsilon}^{p}_{ j,{[l/2]}}\rangle
 \langle \tilde{\epsilon}^{p}_{ j+1,l}\rangle},
\end{equation}
where $p$ is an even integer. The notation [...] in equation (26)
denotes the integer part of the quantity. Because the spatial
range of the cell $(j,[l/2])$ is the same as that of the two cells
$(j+1,l)$ and $(j+1,l+1)$, $C^{p,p}_{j}$ measures the correlation
between the fluctuations on scales $j$ and $j+1$ at the {\it same}
physical area. For Gaussian noise, the variables
$\tilde{\epsilon}_{j,l}$ on scales $j$ and $j+1$ are uncorrelated,
and hence we have $\langle \tilde{\epsilon}^p_{ j,{[l/2]}}
  \tilde{\epsilon}^p_{j+1,l}\rangle =
\langle \tilde{\epsilon}^{p}_{\bf j,{[l/2]}}\rangle
 \langle \tilde{\epsilon}^{p}_{\bf j+1,l}\rangle$, or
$C^{p,p}_{\bf j, j+1}=1$. Therefore, the scale-scale correlation
is effective for drawing a non-Gaussian signal from a Gaussian
background, even when the variance of the noise is comparable to
the signal (Feng et al. 2000; Feng \& Fang 2000).

Figure 12 presents the $C^{p,p}_{ j}$ versus $j$ of the $\delta
T_b$ field with parameters $(T_0, \eta, z)= (1.3\times 10^4 K, 2,
7)$ and $(0.8 \times 10^4 K, 2, 10)$ and $p=2$, 4 and 8. Figure 12
shows once again that the statistical results are less dependent
on the temperature parameters considered. For all cases of $p=2$,
4 and 8, the $\delta T_b$ fields are substantially scale-scale
correlated on scales $j >3$, corresponding to length scale 6.25
$h^{-1}$ Mpc and angular scale 3$^{\prime}$.52 at $z=7$ or
3$^{\prime}$.28 at $z=10$. At scale $j=8$, or 0.20 $h^{-1}$Mpc,
$C^{p,p}_{j}$ is as large as $10^2$, which would be very useful
for distinguishing the signal of $\delta T_b(z_0)$ from noise.

From Figure 12 we see that the scale-scale correlation is
remarkably increasing from $p=2$ to $p=4$, but not so from $p=4$
to $p=8$. This is again because the PDFs of $\delta T_b$ are long
tailed, but not very long. This can strongly affect statistics of
order $n\leq 5$, but not $n > 5$. We calculate also the
scale-scale correlation for noise-contaminated samples. It indeed
shows that the Gaussian noise can be filtered out with the
scale-scale correlations.

\section{Discussion and Conclusion}

We have studied the brightness temperature excess $\delta T_b$ of
the redshifted 21 cm emission of hydrogen clouds in the epoch$z_r
> z > z_{GP}$. According to the LN model, the mean temperature of
hydrogen clouds does not undergo strong evolution and is generally
$T_0\simeq 10^4$ K.  Therefore, the 21 cm emission from this epoch
possesses some statistical features weakly dependent on the
details of this epoch. They are as follows:

1. The random field of $\delta T_b$ is substantially non-Gaussian.
It consists of spikes with high $\delta T_b$ and a low $\delta
T_b$ area between the spikes.

2. The mean of $\delta T_b$ is $\simeq 2 $ mK, and the variance
$\langle (\delta T_b)^2\rangle^{1/2}$ is of the order of 10 mK on
about arcminute scales, while the spikes can be as high as 40 mK.

3. The one-point distributions of either $\delta T_b$ or
$\widetilde{\delta T_b}_{,(j,l)}$ have long tails approximately
following power laws.

4. The $n$th-order moment of $\delta T_b$ is quickly increasing
with $n$ when $n\leq  5$, and slowly when $n>5$. It is very
different from a Gaussian field and an LN field.

5. The scale-scale correlation is significant for all scales,
which is effective for washing out Gaussian noise contamination.

All these results show that although the clustering of hydrogen
gas is suppressed at $z<z_r$, the 21 cm emission fields are
significantly non-Gaussian since $z_r$.

To obtain the above results, we used the fraction of neutral
hydrogen $HI$ given by the LN model (Liu et al. 2004). Actually,
the evolution of the $HI$ fraction in $z_r > z>z_{GP}$ is not
strongly constrained by the current observations. The optical
depth $\tau_e$ is sensitive to $z_r$ and the mean of the $HI$
fraction in $z_r > z>z_{GP}$, but not to the details of the
evolution of the $HI$ fraction. Different evolution models of the
$HI$ and $HII$ fractions can reasonably explain the observed
$\tau_e$. However, among the above results, the last two points
(high-order moments and scale-scale correlations) are based on
statistics given by the ratios of $\delta T_b$. They are probably
less dependent on the $HI$ or $HII$ fractions.

We show that the non-Gaussian properties of the $\delta T_b$ field
comes from the non-Gaussianity of the mass field at high
redshifts. For a given redshift $z$, the $\delta T_b$ field of the
redshifted 21 cm emission is dependent only on $T_m$ at that
redshift, but less sensitive to the details of the redshift
evolution of $T_m$. On the other hand, in the epoch $z_r >
z>z_{GP}$, $T_m$ is in the range $10^3-10^4$ K. Therefore, the
non-Gaussian features revealed in this paper may help in, or even
be necessary for, unambiguously identifying the 21 cm emission
from the epoch $z_r > z>z_{GP}$.

\acknowledgments

P.H. is supported by a Fellowship of the World Laboratory. L.L.F.
acknowledges support from the National Science Foundation of China
(NSFC) and the National Key Basic Research Science Foundation.
This work is partially supported by the National Natural Science
Foundation of China (10025313) and the National Key Basic Research
Science Foundation of China (NKBRSF G19990752).

\newpage

\figcaption {Comoving Jeans length $x_J$ as a function of the
cosmic scale factor $1/(1+z)$. In the period $z_r<z<z_{GP}$, the
mean temperature of hydrogen is assumed to be
$T_m=T_0[(1+z)/(1+z_r)]^{2}$, with $T_0=1.3\times 10^4$ K (solid
line) and 0.8$\times 10^4$ K (dotted line).} \label{Fig1}

\figcaption {Variance $\sigma_0$ of linear perturbations as a
function of $1/(1+z)$, which is calculated with the $x_J$ given by
Fig. 1.} \label{Fig2}

\figcaption {Temperature evolution in the three epochs: (1)
$z>z_r=18$, $T=4[(1+z)/16]^{2}$ K; (2) $z_r > z > z_{GP}=7$,
$T=T_0[(1+z)/(1+z_r)]^{2}$, with $T_0=1.3\times 10^4$ K (solid
line), and 0.8$\times 10^4$ K (dotted line); (3) $z<z_{GP}$, $T
\simeq T_0$. The increase of $T$ at $z=3.3$ is due to $He$
reionization.} \label{Fig3}

\figcaption {Realization of hydrogen density fields at redshifts
7, 10 and 15. The densities are in units of ${\bar\rho}_{IGM}$.}
\label{Fig4}

\figcaption {Realization of the fields of $T_b$, $\tau_{21}$ and
$\delta T_b$. The relevant parameters $(T_0, \eta, z)$ are taken
to be $(1.3\times 10^4$ K, 2, 7).} \label{Fig5}

\figcaption {Cumulative $\langle \delta T_b^2\rangle^{1/2}(>\rho)$
mK for field with density $>\rho$ as a function of $\rho$. The
parameters $(T_0, \eta, z)$ are taken to be $(1.3\times 10^4$ K,
2, 7).} \label{Fig6}

\figcaption {One-point distributions of $\delta T_{b,(j,l)}$ for
$(T_0, \eta, z)= (1.3\times 10^4$ K, 2, 7) (left panels) and
$(0.8\times 10^4$ K, 2, 10) (right panels). The smoothing scales
are $j=6,$ 8, and 10.}\label{Fig7}

\figcaption {One-point distributions of $\widetilde{\delta
T_b}_{,(j,l)}$ for $(T_0, \eta, z)= (1.3\times 10^4$ K, 2, 7)
(left panels) and $(0.8\times 10^4$ K, 2, 10) (right panels). The
smoothing scales are $j=6,$ 8, and 10. } \label{Fig8}

\figcaption {Two-mode correlation function $\ln \langle\delta
T_{b,(j,l)}\delta T_{b,(j,l')}\rangle$ vs. $\ln r$ ($h^{-1}$ Mpc)
for parameters $(T_0, \eta, z)= (1.3\times 10^4$ K, 2, 7). The
smoothing scales are $j=8$, 10, and 12.} \label{Fig9}

\figcaption {DWT power spectrum $P_{j}$ of the $\delta T_b$ field
for parameters $(T_0, \eta, z)= (1.3\times 10^4$ K, 2, 7) and
$(0.8\times 10^4$ K, 2, 10).} \label{Fig10}

\figcaption {The n-dependence of (a) $\langle (\delta T_{b,
j}-\langle\delta T_{b,j}\rangle)^{2n}\rangle^{1/n}/ \langle
(\delta T_{b, j}-\langle\delta T_{b,j}\rangle)^2\rangle$ and (b)
$\langle (\delta T_{b,j}^n \rangle^{1/n}/\langle (\delta
T_b)\rangle$ for parameters $(T_0, \eta, z)= (1.3\times 10^4$ K,
2, 7) (left panels) and $(0.8\times 10^4$ K, 2, 10) (right panels)
} \label{Fig11}

\figcaption {Scale-scale correlation $C^{p,p}_j$ vs. $j$ for
parameters $(T_0, \eta, z)= (1.3\times 10^4$ K, 2, 7) (left
panels) and $(0.8\times 10^4$ K, 2, 10) (right panels); $p$ is
taken to be 2, 4, and 8.} \label{Fig12}

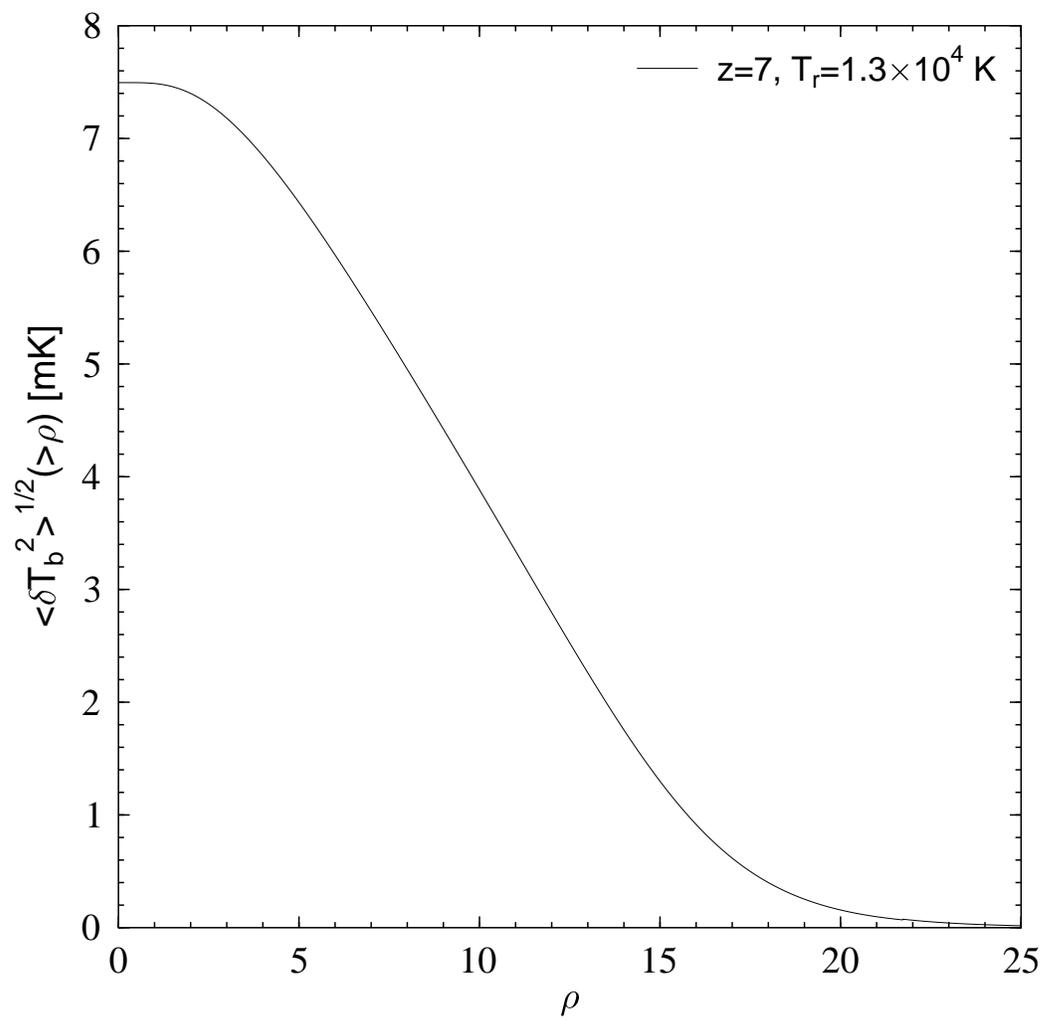
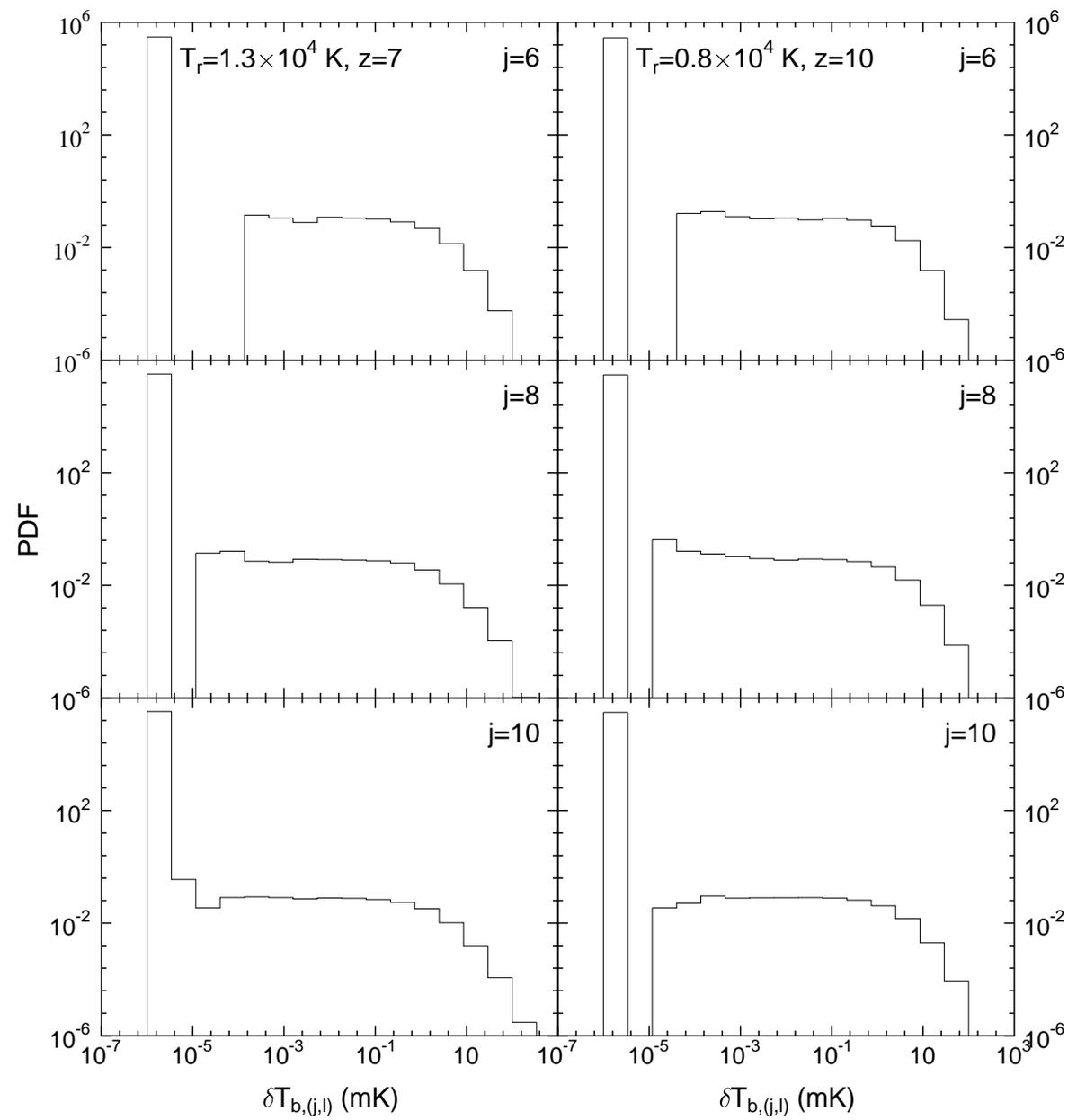
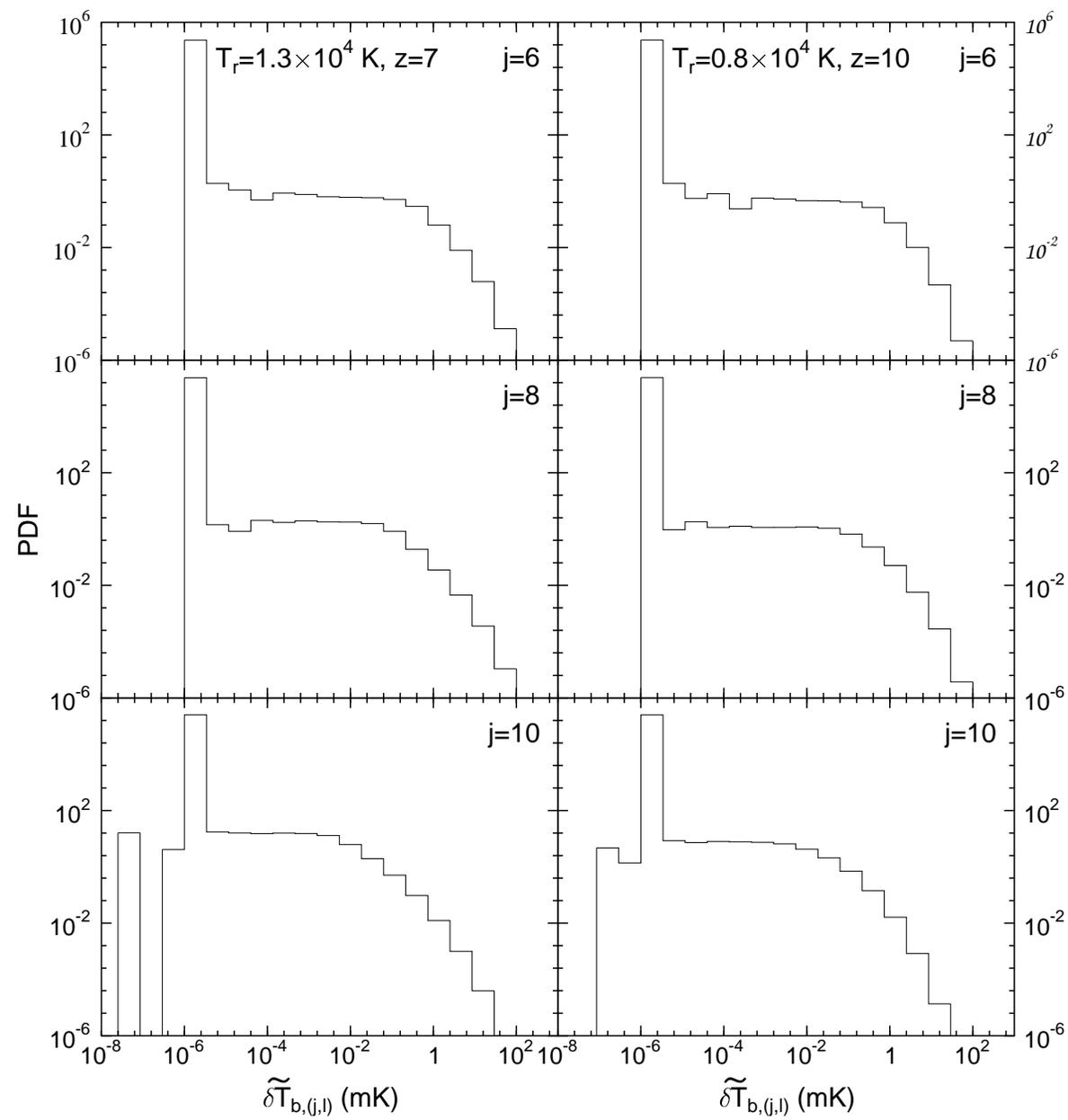
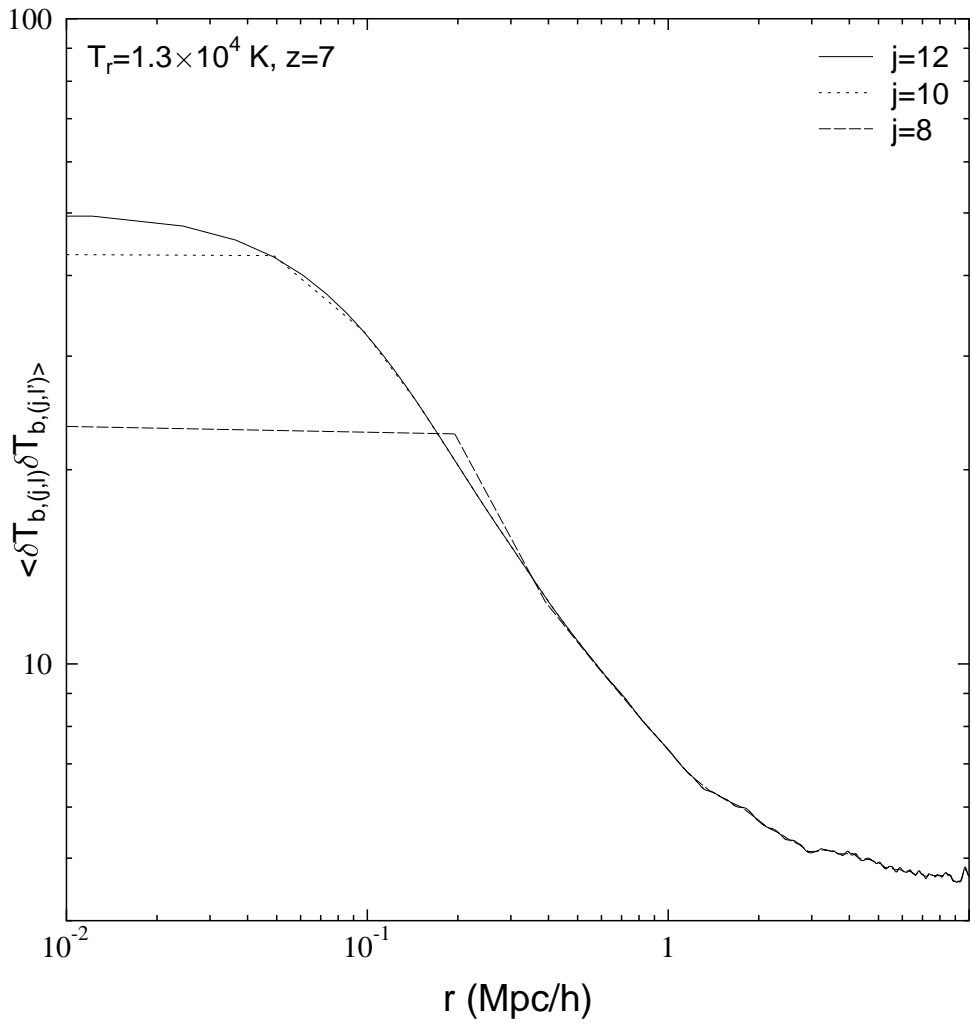
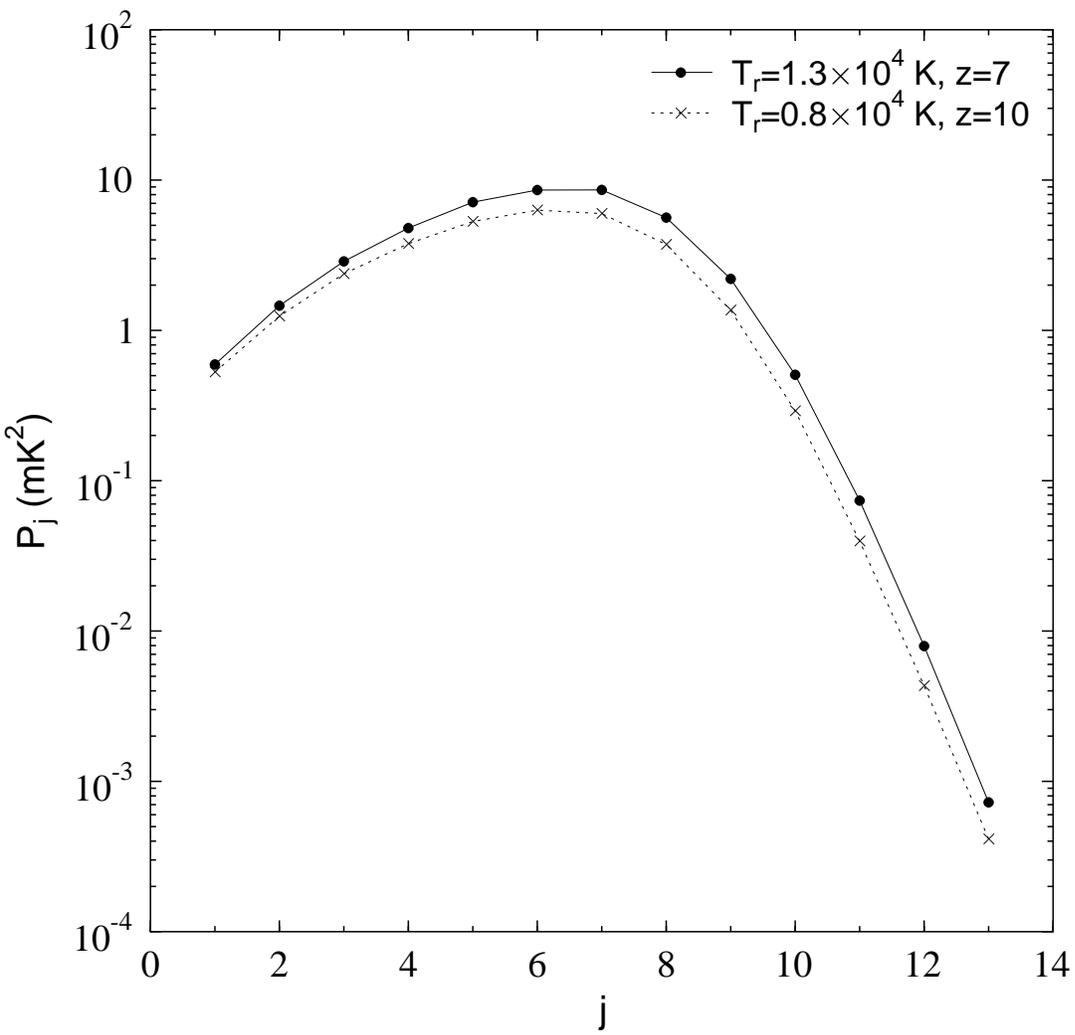
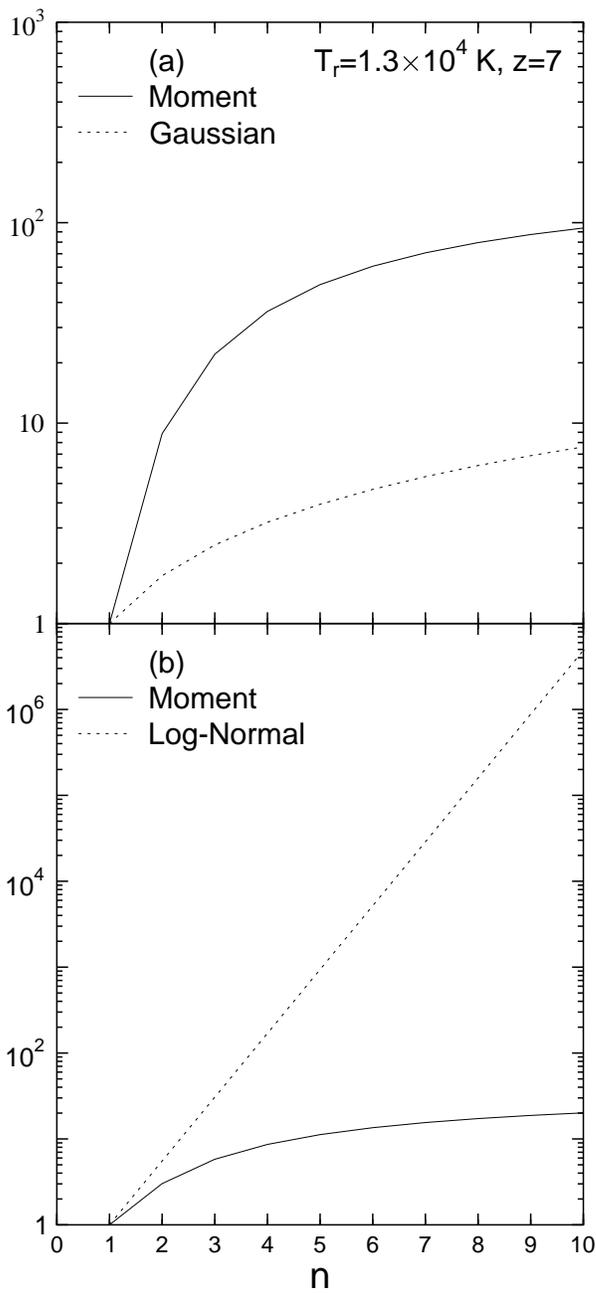
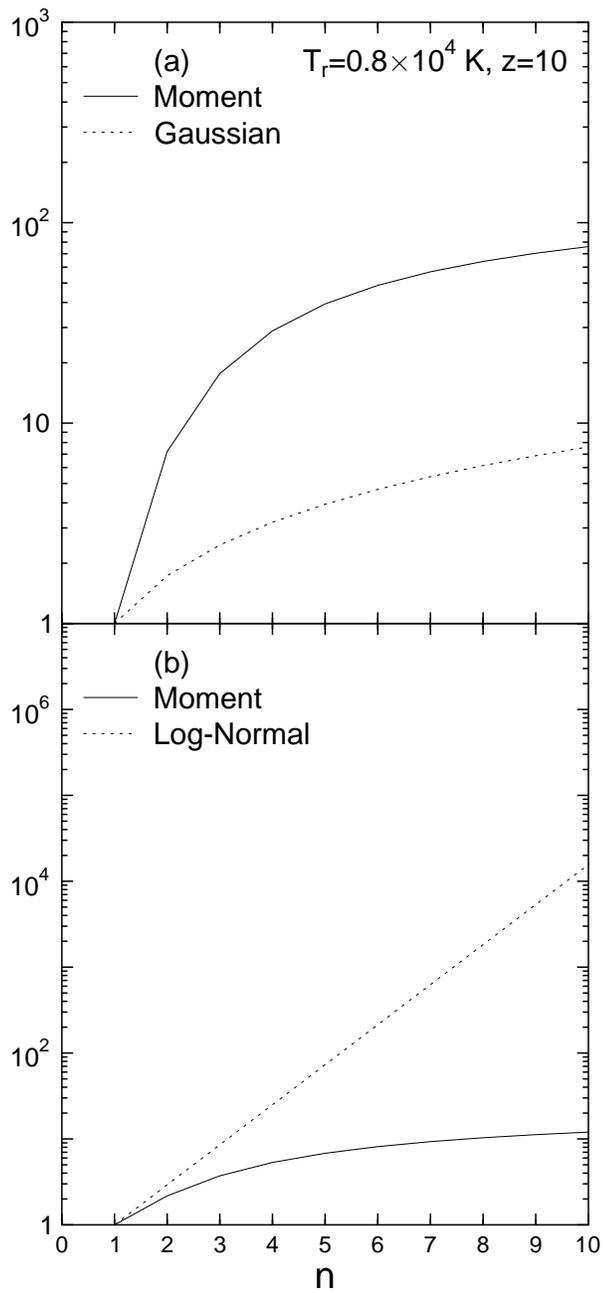
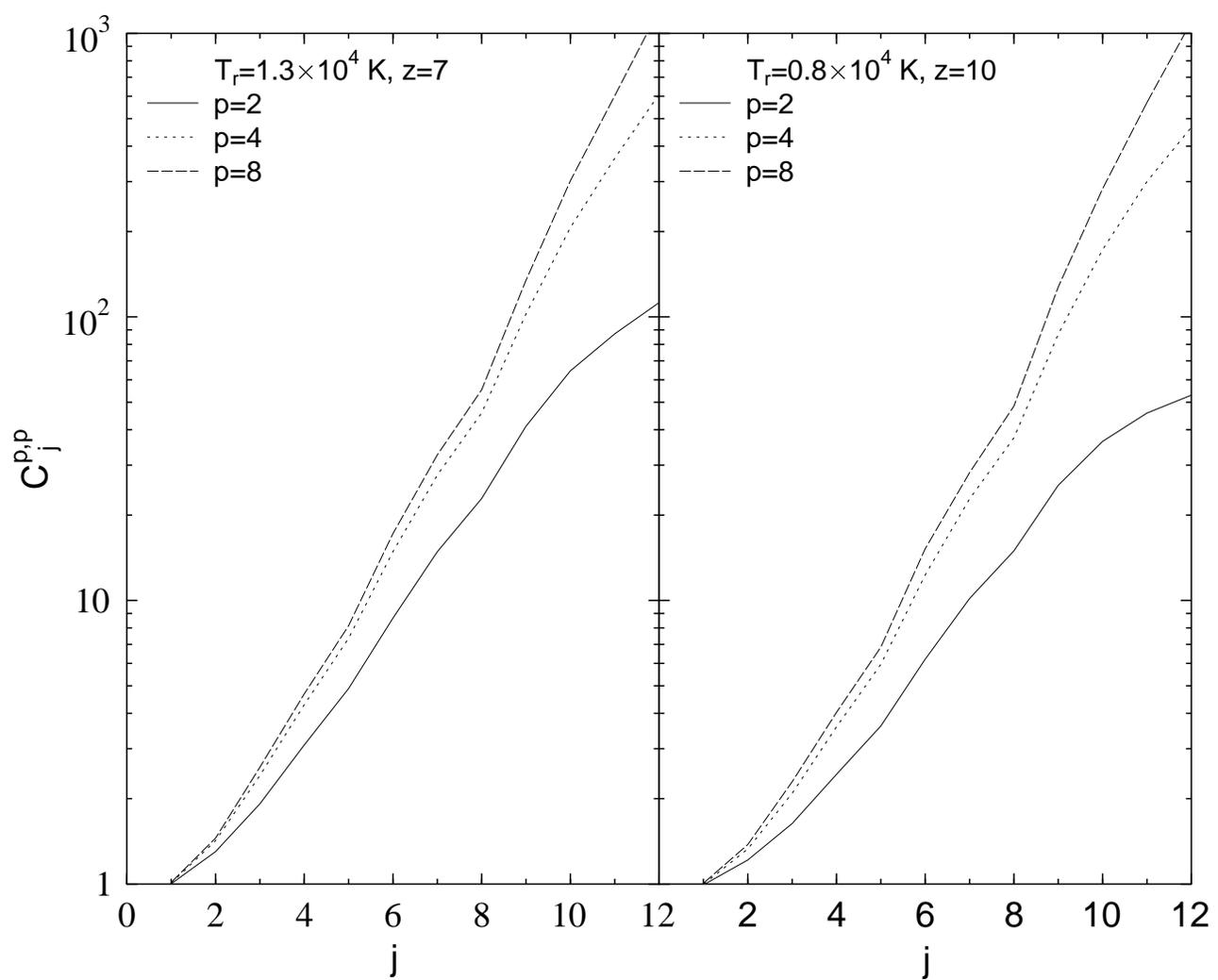
\end{document}